\documentclass{webofc}
\usepackage[varg]{txfonts}   
\usepackage{braket}
\begin{document}
\title{Compositeness of near-threshold exotic hadrons with \\ decay and coupled-channel effects}
%
%


\author{\firstname{Tomona} \lastname{Kinugawa}\inst{1}\fnsep\thanks{\email{kinugawa-tomona@ed.tmu.ac.jp}} \and
        \firstname{Tetsuo} \lastname{Hyodo}\inst{1}\fnsep\thanks{\email{hyodo@tmu.ac.jp}}
}

\institute{Department of Physics, Tokyo Metropolitan University, Hachioji 192-0397, Japan}

\abstract{The near-threshold exotic hadrons such as $T_{cc}$ and $X(3872)$ are naively considered as the hadronic molecular state from the viewpoint of the low-energy universality. However, it is also known that the elementary dominant state is not completely excluded as the internal structure of the near-threshold states. Furthermore, the dominance of molecules is expected to be modified by the decay or coupled channels. We discuss these features of the near-threshold bound states by calculating the compositeness with the effective field theory.}
\maketitle
\section{Introduction}
\label{sec:intro}

The exotic hadrons have been of interest in hadron physics. It is considered that the exotic hadrons have a different internal structure from ordinary hadrons ($qqq$ or $\bar{q}q$), such as the multiquarks or hadronic molecules~\cite{Guo:2017jvc,Brambilla:2019esw}. Recent experiments reported the observation of exotic hadrons, including $X(3872)$ and $T_{cc}$ by the Belle collaboration~\cite{Belle:2003nnu} and by the LHCb collaboration~\cite{LHCb:2021}, respectively. One of the characteristic features of $T_{cc}$ and $X(3872)$ is the fact that these are observed in the near-threshold energy region. 

The low-energy universality is known as an important notion to constrain the property of the near-threshold states~\cite{Braaten:2004rn,Naidon:2016dpf}. The universality holds when the magnitude of the scattering length $|a_{0}|$ is sufficiently larger than other length scales in the system. In this case, the microscopic features of the interaction are irrelevant, and all physical quantities are scaled only by $|a_{0}|$. For example, the binding energy $B$ is given by $B = 1/(2\mu a_{0}^{2})$ with the reduced mass $\mu$. The universality indicates that the binding energy is smaller than other energy scales, because $|a_{0}|$ is much larger than other length scales. As a consequence, the low-energy universality holds when the binding energy is sufficiently small.

The internal structure of the near-threshold states can be quantitatively expressed by the compositeness $X$~\cite{Hyodo:2011qc,Aceti:2012dd}. The compositeness $X$ is schematically written as the weight of the hadronic molecular component $\ket{{\rm molecule}}$ in the bound state $\ket{\psi}$:
\begin{align}
X&=|\braket{{\rm molecule\ }|\ \psi}|^{2}.
\end{align}
For the bound states, $X$ can be regarded as the probability because it is a real value and $0\leq X\leq 1$. Therefore, we can quantitatively classify the bound states whether they are composite dominant ($X>0.5$) or elementary dominant ($X<0.5$). 

In the weak-binding limit $B\to 0$, it is shown that the compositeness $X$ goes to unity from the viewpoint of the low-energy universality~\cite{Hyodo:2014bda,Kinugawa:2022fzn}. In this sense, it is naively expected that the compositeness of the near-threshold states with $B\neq0$ but $B\sim 0$ is close to unity, $X\sim 1$, and the states are composite dominant. However, the state can be elementary dominant with a fine tuning of the parameters, no matter how small the binding energy is~\cite{Hanhart:2014ssa}. In this study, we aim to show the features of the near-threshold bound states from the viewpoint of the fine tuning and the low-energy universality. More detailed discussions can be found in Ref.~\cite{Kinugawa:2023fbf}.

\section{Universality of near-threshold states}
\label{sec:near-threshold}

In this section, we focus on the weakly bound state in the single-channel $\psi_{1}\psi_{2}$ scattering. For this purpose, we introduce the effective field theory model with the following Hamiltonian:
\begin{align}
\mathcal{H}&=\frac{1}{2m_{1}}{\nabla}\psi_{1}^{\dag}\cdot {\nabla}\psi_{1}+\frac{1}{2m_{2}}{\nabla}\psi_{2}^{\dag}\cdot {\nabla}\psi_{2}+\frac{1}{2M}{\nabla}\phi^{\dag}\cdot {\nabla}\phi+\nu_{0}\phi^{\dag}\phi
+g_{0}(\phi^{\dagger}\psi_{1}\psi_{2}+\psi^{\dagger}_{1}\psi^{\dagger}_{2}\phi).
\label{eq:H-wbs}
\end{align}
Here $m_{i}$ ($M$) is the mass of $\psi_{i}$ (the bare state $\phi$), $\nu_{0}$ is the energy of the bare state which is measured from the threshold of the $\psi_{1}\psi_{2}$ scattering, and $g_{0}$ is the coupling constant of the contact three-point interaction. From the Lippmann-Schwinger equation, the amplitude of the $\psi_{1}\psi_{2}$ scattering is obtained as
\begin{align}
f(k)&=-\frac{\mu}{2\pi}\left[\frac{\frac{k^{2}}{2\mu}-\nu_{0}}{g_{0}^{2}}+\frac{\mu}{\pi^{2}}\left[\Lambda+ik\arctan\left(-\frac{\Lambda}{ik}\right)\right]\right]^{-1},
\label{eq:1ch-amplitude}
\end{align}
where $\mu$ is the reduced mass, and the cutoff $\Lambda$ is introduced to avoid the divergence of the momentum integral in the loop function. We consider that the system has a bound state with the binding energy $B$. We can derive the exact expression of the compositeness $X$ of the bound state in this model by applying the method in Ref.~\cite{Kamiya:2016oao}:
\begin{align}
X&=\left[1+\frac{\pi^{2}\kappa}{g_{0}^{2}\mu^{2}}\left(\arctan\left(\frac{\Lambda}{\kappa}\right)-\frac{\frac{\Lambda}{\kappa}}{1+\left(\frac{\Lambda}{\kappa}\right)^{2}}\right)^{-1}\right]^{-1}, \quad \kappa=\sqrt{2\mu B}.
\label{eq:X-1ch}
\end{align}

There are three model parameters, the coupling constant $g_{0}$, bare state energy $\nu_{0}$, and cutoff $\Lambda$. Here we fix the binding energy $B$ to discuss the compositeness of the weakly bound state. From the bound state condition $f^{-1}(i\kappa)=0$, the square of the coupling constant $g_{0}^{2}$ is written by $B$, $\nu_{0}$, and $\Lambda$:
\begin{align}
g_{0}^{2}(B;\nu_{0},\Lambda)&=\frac{\pi^{2}}{\mu}(B+\nu_{0})\left[\Lambda-\kappa\arctan\left(\frac{\Lambda}{\kappa}\right)\right]^{-1}.
\label{eq:g02}
\end{align}
We can further eliminate one more degree of freedom by using the dimensionless quantities by $\Lambda$. Eventually, the energy of the bare state $\nu_{0}$ remains as a parameter. In general, the compositeness $X$ has the model dependence because it is not an observable. The choice of the value of $\nu_{0}$ represents the choice of the model. Therefore, we can investigate the model dependence of $X$ by examining its $\nu_{0}$ dependence. In this work, we vary $\nu_{0}$ in the region $-B\leq \nu_{0}\leq E_{\rm typ}$, with $E_{\rm typ}=\Lambda^{2}/(2\mu)$ as the typical energy scale in the model. The lower boundary of $\nu_{0}$ region is determined to have a real coupling constant ($g_{0}^{2}\geq 0$), and the upper is given not to exceed the applicable limit of the effective field theory.

\begin{figure}[tb]
\centering
\includegraphics[width=6cm,clip]{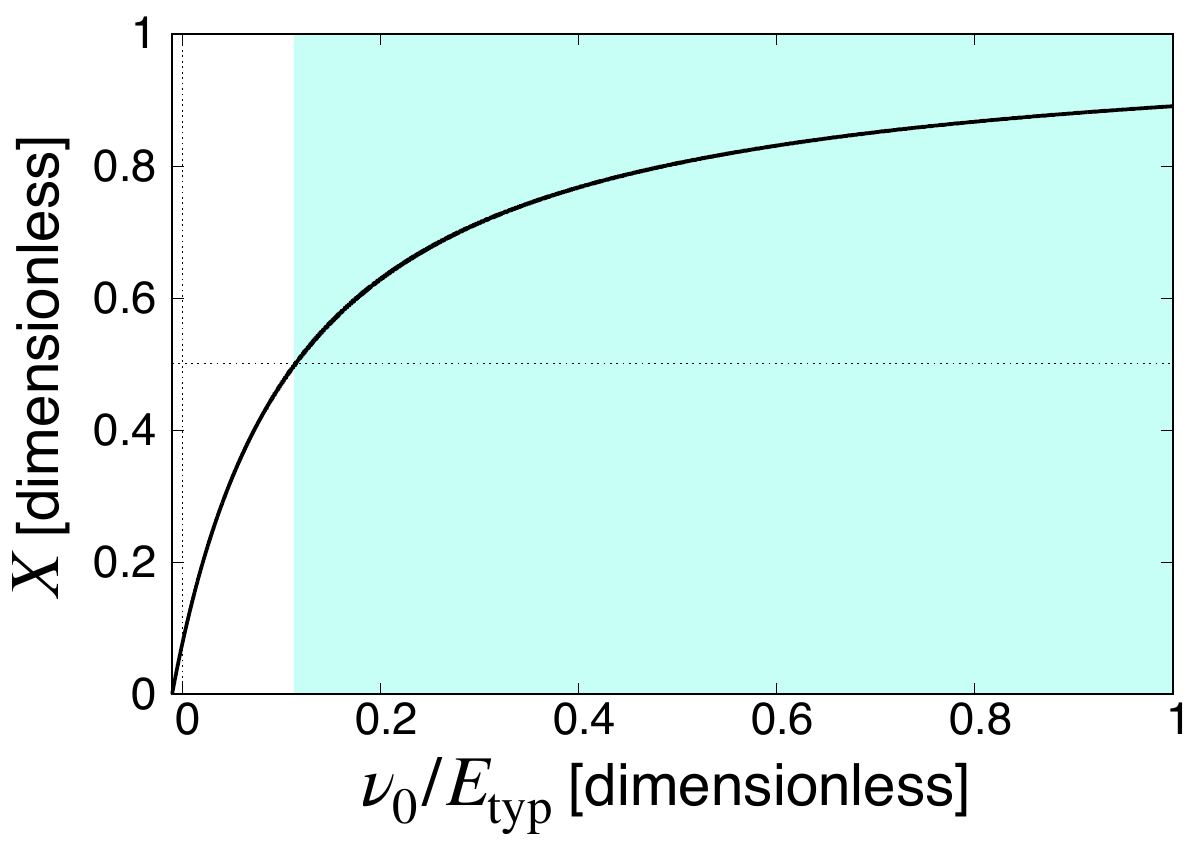}
\caption{The compositeness $X$ of the weakly bound state ($B=0.01E_{\rm typ}$) as a function of the model parameter $\nu_{0}$. The shaded region is the parameter region with $X(\nu_{0})>0.5$.}
\label{fig:near-threhsold}      
\end{figure}

In Fig.~\ref{fig:near-threhsold}, we plot the compositeness $X$ of the weakly bound state with $B=0.01E_{\rm typ}$ as a function of the model parameter $\nu_{0}$. We see that $X>0.5$ for 88~\% of the $\nu_{0}$ region (shaded region). From this result, it is quantitatively shown that the weakly bound state is composite dominant in most of the parameter region, and the realization of the low-energy universality is demonstrated with a simple model. At the same time, even for the weakly bound state, we show that the elementary dominant state with $X<0.5$ can be realized with a fine tuning of $\nu_{0}$ to 12~\% of the parameter region.

\section{Application to $T_{cc}$ and $X(3872)$}
\label{sec:application}

Most of the exotic hadrons have couplings to other channels in addition to the threshold channel. In this case, the universal nature of the near-threshold states shown in the previous section can be modified. Therefore, in this section, we calculate the compositeness of the actual exotic hadrons $T_{cc}$ and $X(3872)$ with the effects of the decay and coupled channels, and discuss the modification of the universality implications by channel couplings. We use the complex coupling constant to effectively take into account the decay contribution, and introduce the coupling to the isospin partner channel above the threshold~\cite{Kinugawa:2023fbf}. In this case, $\tilde{X}_{1}$ stands for the compositeness of the threshold channel, and $\tilde{X}_{2}$ for that of the coupled channel~\cite{Sekihara:2015gvw,Kinugawa:2023fbf}. For the calculation, we use the central value of the pole position in Ref.~\cite{LHCb:2021} for the eigenenergy of $T_{cc}$ ($E=-0.36-0.024i$ MeV), and the value in Particle Data Group~\cite{ParticleDataGroup:2022pth} for the eigenenergy of $X(3872)$ ($E=-0.04-0.595i$ MeV). We employ the cutoff $\Lambda=m_{\pi}=140$ MeV by assuming the $\pi$ exchange as the long-range interaction between $D$ mesons.

\begin{figure}[tb]
\centering
\includegraphics[width=6cm,clip]{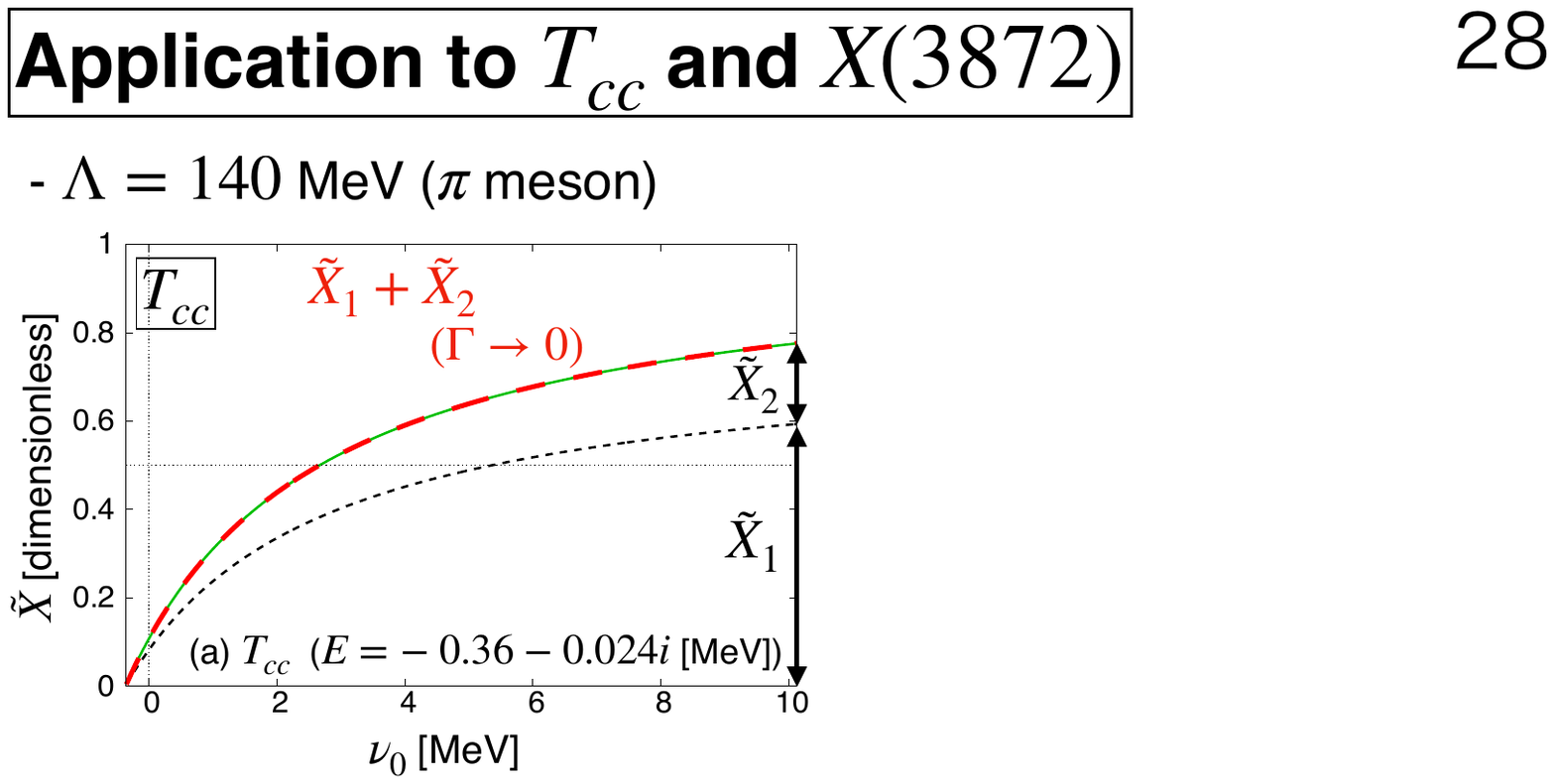} 
\includegraphics[width=6cm,clip]{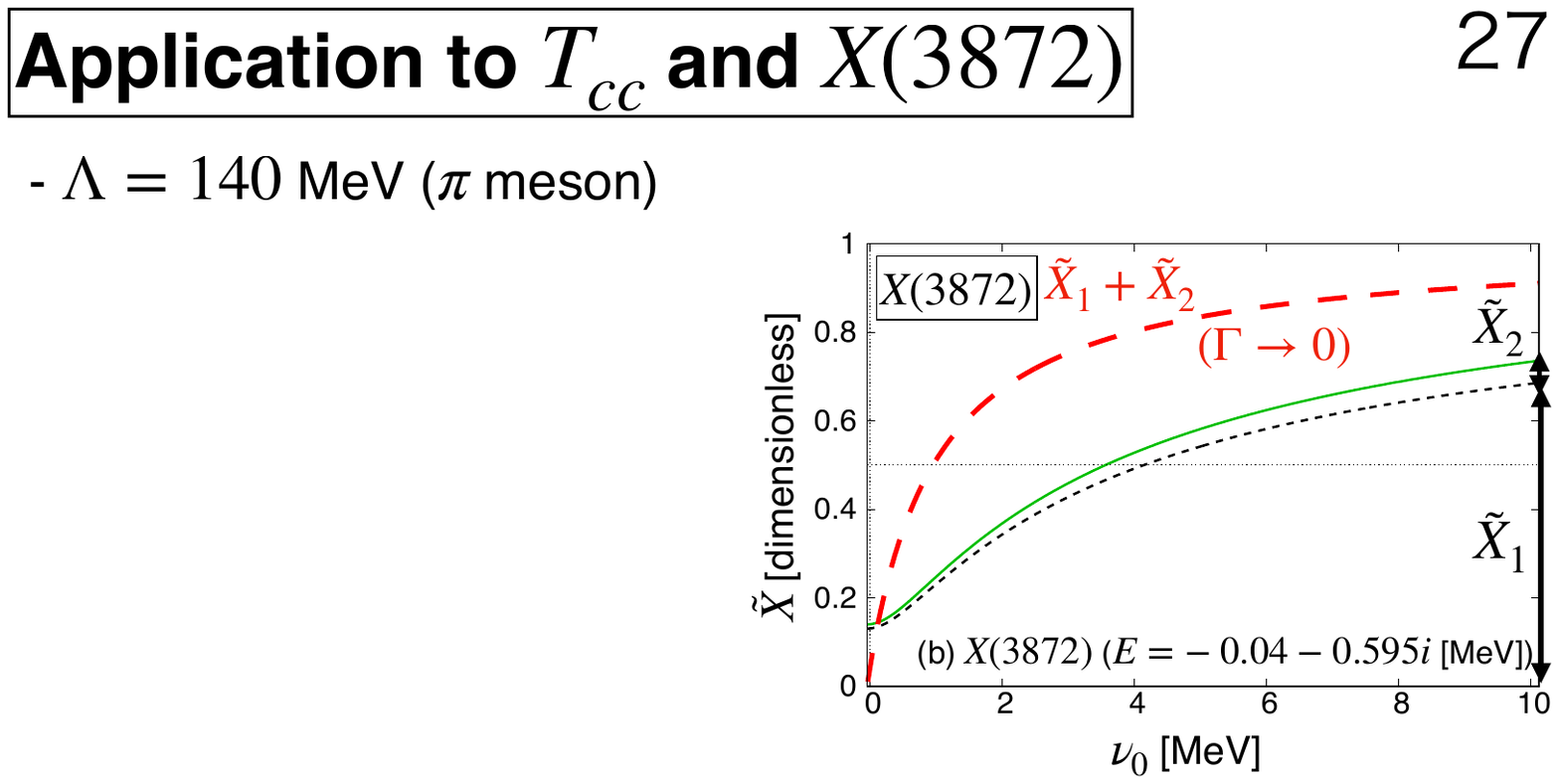} 
\caption{The compositeness of $T_{cc}$ (left panel) and $X(3872)$ (right panel) as a function of the model parameter $\nu_{0}$. The dotted lines correspond to the compositeness of the threshold channel $\tilde{X}_{1}$, the solid lines to the sum of the compositeness of threshold and coupled channels $\tilde{X}_{1}+\tilde{X}_{2}$, and the dashed lines to the sum of the compositeness in the bound state limit $\Gamma\to 0$.}
\label{fig:apply}      
\end{figure}

In Fig.~\ref{fig:apply}, we show the $\nu_{0}$ dependence of the compositeness of $T_{cc}$ and $X(3872)$. The dotted lines stand for the compositeness of the threshold channel $\tilde{X}_{1}$, solid lines for the sum of the compositeness of threshold and coupled channels $\tilde{X}_{1}+\tilde{X}_{2}$. To see the decay contribution, $\tilde{X}_{1}+\tilde{X}_{2}$ in the bound state limit $\Gamma\to0$ is plotted by the dashed lines ($\Gamma$ is the decay width). In the left panel, we see that there is almost no difference between the solid and dashed lines. Therefore, the compositeness of $T_{cc}$ is not affected very much by the decay contribution. This reflects the much smaller decay width of $T_{cc}$ ($\Gamma=0.048$ MeV) than the binding energy ($B=0.36$ MeV). In contrast, for $X(3872)$, the difference between the solid and dashed lines is sizable. This is because of the larger decay width of $X(3872)$ ($\Gamma=1.19$ MeV) than the binding energy ($B=0.04$ MeV). By comparing $\tilde{X}_{2}$ of $T_{cc}$ with that of $X(3872)$, the effect of the coupled channel contributes more to the compositeness of $T_{cc}$. The reason is that the threshold energy difference of $T_{cc}$ (1.41 MeV) is smaller than that of $X(3872)$ (8.23 MeV). In both cases, we quantitatively show that the threshold-channel compositeness $\tilde{X}_{1}$ decreases by introducing the decay and coupled channel effects. Furthermore, we find that $\tilde{X}_{1}$ largely decreases with the larger decay width or the smaller threshold energy difference. 

\section{Summary}
\label{sec:summary}
In this study, we have studied the nature of the weakly-bound states with the compositeness. At first, in the single-channel scattering model, we have quantitatively demonstrated that the elementary dominant state is always realized even for the weakly bound states but with a significant fine tuning. Then, the decay effect and coupled channel are introduced to calculate the compositeness of $T_{cc}$ and $X(3872)$. We have shown the importance of the coupled channel (decay effect) for $T_{cc}$ [$X(3872)$] with the small threshold energy difference (large decay width) to consider their compositeness.

\end{document}